\documentstyle[twoside,fleqn,espcrc2,epsfig]{article}

\newcommand{\la}[1]{\label{#1}}

\newlength{\numlen}
\newcommand{\n}{\settowidth{\numlen}{0}\makebox[\numlen]{}}
\newcommand{\cen}[1]{\multicolumn{1}{c}{#1}}
\newlength{\indeksinpituus}

\newlength{\ypit}

\newcommand{\be}{\begin{equation}}
\newcommand{\ee}{\end{equation}}
\newcommand{\ba}{\begin{eqnarray}}
\newcommand{\ea}{\end{eqnarray}}

\newcommand{\etal}{{et al.\ }}
\newcommand{\eq}{eq.~}

\newcommand{\fig}{fig.~}

\newcommand{\nr}[1]{(\ref{#1})}

\newcommand{\oo}{{\rm o}}

\newcommand{\den}{n}
% ...

\newcommand{\AmS}{{\protect\the\textfont2
  A\kern-.1667em\lower.5ex\hbox{M}\kern-.125emS}}

\title{Multicanonical Cluster Algorithm}

\author{K. Rummukainen\address{CERN/TH, CH-1211 Geneva 23, Switzerland}}

\begin{document}

\begin{abstract}

In this talk I present a multicanonical hybrid-like two-step
algorithm, which consists of a microcanonical spin system update with
demons, and a multicanonical demon refresh.  The demons act as a
buffer between the multicanonical heat bath and the spin system,
allowing for a large variety of update schemes.  In this work the
cluster algorithm is demonstrated with the 2-dimensional 7-state Potts
model, using volumes up to $128^2$.
\end{abstract}

\maketitle

\section{Introduction}

\begin{figure}[bt]
\vspace{0.9cm}
\hfill\epsfig{height=5.5cm,file=fig1.ps}\hspace{2mm}\mbox{}
\vspace{0.0cm}
\caption[0]{The probability distributions $p_{\beta_c}(E)$ of the 2d7s
Potts model.}
\label{fig1}
\end{figure}

Multicanonical methods \cite{Berg91} provide a way to effectively
study ``difficult'' regions of the phase space, the most
familiar example perhaps being the mixed phase of a system with a first-order
phase transition.  In a thermally driven order-disorder transition,
the probability distribution $p(E)$ in a finite volume $V=L^2$ near
the transition temperature develops a double-peak structure
(\fig\ref{fig1}).  The probability of the mixed phase configurations
is suppressed by an amount $\sim \exp -\sigma A$, where $\sigma$ is
the interface tension between the coexisting pure phases, and $A$ is
the total area of the interface.  In a multicanonical simulation this
suppression is compensated by explicitly enhancing the probability of
the suppressed configurations.  Usually this is achieved by
substituting the Boltzmann weight $\exp -\beta E$, where $\beta=1/T$,
with a multicanonical weight $\exp -W(E)$.  The function $W(E)$ is
carefully chosen to boost the probability of the mixed states with
energies between the bulk phase values $E_{\oo}$ and $E_{\rm do}$.
The non-linearity of $W(E)$ makes the update effectively non-local,
thus preventing cluster update algorithms and making
vectorization difficult.

In this talk I present a two-step multicanonical algorithm: first, a
spin system is updated {\em microcanonically\,} \cite{Creutz83} with a
set of demons, and secondly, the demons are refreshed with a {\em
multicanonical\,} heat bath.  The demons act as a buffer, isolating
the actual system from the multicanonical heat bath, thus enabling one
to choose an optimal microcanonical update step for a particular
problem.  Here, a microcanonical cluster algorithm \cite{Creutz92} is
applied to the 2-dimensional 7-state Potts model, defined by the
partition function \cite{Potts52}
\begin{eqnarray}
Z(\beta)& = &\mbox{$ \sum_{\{s\}} \exp[-\beta E(s)]$} \\
E(s)& =& \mbox{$\sum_{(i,j)} (1-\delta(s_i,s_j))$}, \, s_i = 1\ldots 7 \, .
\end{eqnarray}
This model has already been studied with the standard multicanonical
method \cite{Janke92}.  The simulations are listed in table
\ref{table1}; for the $L^2 = 128^2$ lattice two separate runs were performed.
A more detailed report of this work is published in
ref.~\cite{Rummukainen92}.

\begin{table}
\center
\caption[1]{The number of iterations, the tunnelling time
 and the interface tension from the
2d7s Potts model simulations.\la{table1}}
\begin{tabular}{lrrll}\hline
\cen{$L$} & iterations &\cen{$\tau_L$}& \cen{$\sigma$}&\\ \hline
{\n}20 & 2 500 000 &320(5)\n\n&0.0189(3)   \\
{\n}32 & 5 000 000 &821(15)\n &0.0169(2)   \\
{\n}64 & 6 000 000 &2700(81)\n&0.0147(4)   \\
$128_a$ & 9 000 000&10720(520)&0.01302(17) \\
$128_b$ & 6 000 000&10520(620)&0.01306(21) \\
\hline
\end{tabular}
\end{table}

\section{The Multicanonical Distribution}

Let us now connect the spins and the demons to the multicanonical heat
bath.  The probability distribution is, as a function of the spin system
energy $E_S$ and the demon energy $E_D$,
\be
p(E_S,E_D) \propto \den_S(E_S)\den_D(E_D)\,e^{-G(E_T)} \, ,
\la{mdprob}
\ee
where $\den_S(E_S)$ and $\den_D(E_D)$ are the (unknown) spin system
and the (known) demon density of states, respectively; $E_T=E_S+E_D$
is the total energy, and $G(E_T)$ is the multicanonical weight
function.  In the most general case the weight $G$ is a function of
both $E_S$ and $E_T$, but when $E_T$ is fixed, $p(E_S)$ has to reduce
to the microcanonical distribution $\den_S(E_S)\den_D(E_T-E_S)$,
implying that $G$ can only be a function of $E_T$.  When $E_S$ is
fixed, $p(E_D)$ reduces to a multicanonical distribution for the
demons: $\den_D(E_D)\,e^{-G(E_S+E_D)}$.  Thus the probability
distribution of \eq\nr{mdprob} is preserved in a generic two-step process
of a microcanonical spin update and a multicanonical demon update,
provided that both of the update steps separately satisfy detailed
balance.  Note that if $G(E)=\beta E$, both the demon and the spin
system will have canonical distributions.

The Monte Carlo simulation produces a sample of the distribution of
\eq\nr{mdprob}, from which $\den_S$ can be estimated.
The demon density of states, with $N_D$ demons with discrete energy
values $0,1,2\ldots$, is
\be
\den_D(E_D) = \frac{(N_D - 1 + E_D)!}{(N_D-1)!\,E_D!} \, .
\la{demden}
\ee
The optimal way to solve for $\den_S$ is simply to sum over $E_D$ in
\eq\nr{mdprob} \cite{Rummukainen92}.  This gives
\be
\den_S(E_S) \propto  \frac{\sum_{E_D} p(E_S,E_D)}
	{\sum_{E_D} \den_D(E_D)\,e^{-G(E_T)}}\, .
\la{denstate}
\ee
Note that the result is independent of $E_D$; it is sufficient to measure
only the spin system energy $E_S$.  From $\den_S(E_S)$ one
can calculate various thermodynamical quantities of interest.

In the simulations described here the weight function $G$ was chosen
so as to render the probability distribution as a function of $E_T$ as
flat as possible.  From \eq\nr{mdprob}, one sees that optimally
$G(E_T)\propto\log\den_T(E_T)$, which is a priori unknown.  For the
lattice volumes $\le 64^2$, canonical simulations ($G(E) = \beta E$)
were used to obtain an estimate of $G$, and for the $128^2$ lattice,
finite-size scaling was used to scale up the function used in the
$64^2$ simulation; in this case two runs were performed using slightly
different weight functions.

\section{The Update Algorithm}

The microcanonical spin system + demon update was performed with the
cluster algorithm presented by M.~Creutz \cite{Creutz92}: each {\em
link} on the lattice has a demon, and the cluster is grown across the
link only if (i) the spins at each end of the link have the same value
and (ii) the demon on the link does not have any energy.  This
approach requires $2\times L^2$ demons on a 2-dimensional lattice.
The clusters are flipped to random spin values, and the demon energies
are updated.  After each update cycle the demon locations are
shuffled.  The actual cluster search was performed with the
Hoshen-Kopelman algorithm.

The multicanonical demon refresh was performed after each cluster
sweep.  The main feature of the demon update is that the ``slow'' part
of it can be performed in $\propto\sqrt{V}$ steps.  Here one utilizes
the fact that the demon density of states is known:
first, the new demon energy $E_D^{\rm new}$ is calculated with a {\em
global} heat bath by using the distribution
\be
  p_D(E_D) \propto \den_D(E_D)e^{-G(E_D+E_S)}\,.
\la{heatbath}
\ee
Starting from the old demon configuration, one can add or subtract its
energy until $E_D^{\rm new}$ is reached.  Because the demon energy
fluctuates only $\propto\sqrt{N_D}$, only as many additions or
subtractions are needed.  However, care has to be taken to ensure
proper counting of states: every time a unit of energy is added or
subtracted, a new demon is chosen according to the respective
probabilities \cite{Rummukainen92}
\begin{eqnarray}
p^i_+ &=&  (E_D^i + 1)/(N_D+E_D)
\la{demonadd}\\
p^i_- &=&  E_D^i/E_D  \, ,
\la{demonsub}
\end{eqnarray}
where $E_D^i$ is the energy of the demon $i$.
It can be shown that, by applying $p_+$ to the state $E_D=0$ $k$ times,
one generates all the states with $E_D=k$ with equal probability.

Although the demon refresh described above has only $\propto\sqrt{V}$
steps, an effective implementation of this may require some fast
$\propto V$ operations (for details, see \cite{Rummukainen92}).  On a
Cray X-MP, one full cycle took $<3.9 \mu$s per spin, $\sim 94\%$ of
this was taken by the cluster operations, the rest going to the demon
refresh and the measurements.

\section{Results and conclusions}

\begin{figure}[bt]
\vspace{0.9cm}
\hfill\epsfig{height=5.5cm,file=fig2.ps}\hspace{4mm}\mbox{}
\vspace{0.0cm}
\caption[x]{The tunnelling time as a function of the lattice size.
(a): this work;
(b): multicanonical simulation by Janke~\etal \cite{Janke92};
(c): canonical simulation by Billoire~\etal \cite{Billoire92}.}
\label{fig2}
\end{figure}

The performance of the algorithm is characterized by the tunnelling
time $\tau_L$, which is defined here as one fourth of the average
number of update sweeps that the system needs in order to get from $E_S =
E_o$ to $E_S = E_{do}$ and back; here the energy values $E_o$ and
$E_{do}$ were determined from the locations of the maxima of the
probability distribution at the temperature where the peaks have equal
height (\fig\ref{fig1}).  The times are listed in table~\ref{table1},
and shown in \fig\ref{fig2}.  A power-law fit to the three largest
volumes gives $\tau_L = 1.49(17)\times L^{1.82(3)}$.  Because the
energy gap increases $\propto L^2$, but the average change in the
system energy during one sweep is $\propto L$, $\tau_L$ should
increase as $\mbox{(gap/step)}^2 \propto L^2$.  The observed too small
value of the exponent can be attributed to the shift of the
distribution peak locations as a function of the volume (see
\fig\ref{fig1}).  For comparison, the result from the standard multicanonical
calculation by Janke \etal \cite{Janke92} is $\tau_L = 0.082(17)\times
L^{2.65(5)}$.  In the canonical simulations $\tau_L$ is expected to
grow exponentially as $e^{2\sigma L}$, where $\sigma$ is the interface
tension.  A function of the form $\tau_L = aL^\alpha\,e^{2\sigma L}$
was fitted to the autocorrelation times measured by A.~Billoire
\etal \cite{Billoire92} from lattices up to $64^2$.  The values from the
fit were $a=1.01(15)$ and $\alpha=2.31(4)$, with $\sigma = 0.01174$
kept fixed (see the next paragraph).

The interface tension was measured from the probability distribution
$p_L$ with the Binder method~\cite{Binder82}:
\be
\sigma = - \lim_{L\rightarrow\infty} \log p_L^{\min}/(2L)\,,
\la{tension}
\ee
where the maximum of $p_L$ is normalized to 1 and $p_L^{\min}$ is the
minimum between the peaks at a temperature where the peaks have equal
height (\fig\ref{fig1}).  The minima and maxima of the distributions
were found by fitting a section of parabola close to the extrema; the
results depend on the fitting range only very weakly.  The measured
values of $\sigma$ are listed in table \ref{table1} and shown in
\fig\ref{fig3}, together with the measurements of ref.~\cite{Janke92}.
Using a common finite-size ansatz $\sigma_L = \sigma + c/L$, we obtain
the infinite volume extrapolation $\sigma = 0.01174(19)$.  The result
agrees well with ref.~\cite{Janke92} ($\sigma = 0.0121(5)$), but is
seven standard deviations off from the rigorous value $\sigma =
0.010396\ldots$, recently calculated by Borgs and
Janke~\cite{Borgs92b}.  The errors cited are only statistical, and the
discrepancy is allegedly due to the interactions between the two
interfaces, which are ignored by the finite-size ansatz used.
Still, this result can be compared with the older MC
results~\cite{Kajantie89}, which were $\sim 6$ times too large.

\begin{figure}[bt]
\vspace{0.9cm}
\hfill\epsfig{height=5.5cm,file=fig3.ps}\hspace{4mm}\mbox{}
\vspace{0.0cm}
\caption[0]{The interface tension from this work and from
ref.~\cite{Janke92}.  The line is a fit to the 3 largest volumes and
the arrow is the rigorous infinite volume value \cite{Borgs92b}.}
\label{fig3}
\end{figure}

\begin{figure}[bt]
\vspace{0.9cm}
\hfill\epsfig{height=5.5cm,file=fig4.ps}\hspace{4mm}\mbox{}
\vspace{0.0cm}
\caption[0]{
The transition temperature determined by the equal weight method, and
the finite-size fit to the data.  The dashed line is the exact
infinite-volume $\beta_c$.\la{fig4}}
\end{figure}

As an example of other measurements, we determine the infinite volume
transition temperature with the ``equal weight''
method~\cite{Borgs92}, \fig\ref{fig4}.  A finite-size fit of the form
$\beta_L = \beta_\infty + a\,e^{-bL}$ was fitted to the data, with the
result $\beta_\infty = 1.293562(14)$, which agrees very well with the
exact value $\log(1+\sqrt{7})$.

As a conclusion, we note that the hybrid algorithm presented in this
talk provides an efficient method for performing multicanonical
simulations, at least of discrete spin models.  Generally, the
microcanonical sweep can be faster than even the corresponding
canonical update, and the multicanonical demon refresh can be
performed with little extra effort.  This method can also be
generalized to magnetic transitions by using demons carrying
magnetization.

\end{document}